\documentclass{svjour2}                    % onecolumn
\smartqed  % flush right qed marks, e.g. at end of proof
\usepackage{color}
\usepackage{graphicx}
\begin{document}

%\title{Experience-guided rumor spreading models%\thanks{Grants or other notes
%about the article that should go on the front page should be
%placed here. General acknowledgments should be placed at the end of the article.}
%}
\title{Emergence of influential spreaders in modified rumor models}
%\subtitle{Heterogeneous activity patterns}

%\titlerunning{Experience-guided rumor spreading models}        % if too long for running head
\titlerunning{Rumor dynamics and influential spreaders}

\author{Javier Borge-Holthoefer         \and
        Sandro Meloni         \and
        Bruno Gon\c{c}alves         \and
        Yamir Moreno
}

%\authorrunning{Short form of author list} % if too long for running head

\institute{J. Borge-Holthoefer \at
              Instituto de Biocomputaci\'on y F\'\i sica de Sistemas 
Complejos (BIFI), Universidad de Zaragoza, Mariano Esquillor s/n, 50018 Zaragoza, Spain \\
              \email{borge.holthoefer@gmail.com}
           \and
           S. Meloni \at
              Instituto de Biocomputaci\'on y F\'\i sica de Sistemas 
Complejos (BIFI), Universidad de Zaragoza, Mariano Esquillor s/n, 50018 Zaragoza, Spain
           \and
           B. Gon\c{c}alves \at
              Department of Physics, College of Computer and Information Sciences and
Department of Health Sciences, Northeastern University, Boston MA 02115 USA\\
Aix Marseille UniversitŽ, CNRS UMR 7332, CPT, 13288 Marseille, France
           \and
           Y. Moreno \at
              Instituto de Biocomputaci\'on y F\'\i sica de Sistemas 
Complejos (BIFI), Universidad de Zaragoza, 50018 Zaragoza, Spain\\
Departamento de F\'isica Te\'orica,  Universidad de Zaragoza, 50009 Zaragoza, Spain,\\
Complex Networks and Systems Lagrange Lab, Institute for
   Scientific Interchange, Torino, Italy\\
 }

\date{Received: date / Accepted: date}
% The correct dates will be entered by the editor

\maketitle

\begin{abstract}
The burst in the use of online social networks over the last decade has provided evidence that current rumor spreading models miss some fundamental ingredients in order to reproduce how information is disseminated. In particular, recent literature has revealed that these models fail to reproduce the fact that some nodes in a network have an influential role when it comes to spread a piece of information. In this work, we introduce two mechanisms with the aim of filling the gap between theoretical and experimental results. The first model introduces the assumption that spreaders are not always active whereas the second model considers the possibility that an ignorant is not interested in spreading the rumor. In both cases, results from numerical simulations show a higher adhesion to real data than classical rumor spreading models. Our results shed some light on the mechanisms underlying the spreading of information and ideas in large social systems and pave the way for more realistic diffusion models.
% \PACS{PACS code1 \and PACS code2 \and more}
% \subclass{MSC code1 \and MSC code2 \and more}
\end{abstract}

%%%%%%%%%%%%%%%%%%%%%%%%%%%%%%%%%%%%%%%%%%%%%%
%%%%%%%%%%%%%%%%%%%%%%%%%%%%%%%%%%%%%%%%%%%%%%
\section{Introduction}
\label{intro}
Understanding the way in which a disease or a piece of information spreads from person to person is of obvious practical relevance. If we are able to comprehend the mechanisms that dominate such spreading processes we would be able to enhance the spread of valuable information through a community or impair an outbreak of an infectious disease. The similitudes between these two processes, epidemics and information diffusion (also referred to as rumor spreading \cite{goffman64-1,dietz67-1,galam,kimmel,kosfeld}) have long been recognized and the two fields have evolved in parallel freely borrowing ideas and concepts from each other \cite{DK1,book1_epidemic}. 

In epidemics it has become clear that some individuals that are ``superspreaders'' play a dominant role in the course of an epidemic \cite{smith05-1}. Intuitively, one expects that similarly influential individuals would also be present in the case of information diffusion and recent years have witnessed a growing interest to understand how to identify them~\cite{boccaletti06,castellano2009statistical,watts07-1,rogers03-1,bakshy12-1}. Successful approaches have focused on studying the effect that different network-based centrality measures have on rumor spreading. In particular, one recent seminal approach  \cite{kitsak2010identification} has identified the $k$-core as the best measure to predict influence, outperforming degree centrality or betweenness in the context of an epidemic spreading process. This insight has been followed by many other works, which mainly discuss under which circumstances the $k$-core actually predicts a node's disease spreading capabilities \cite{castellano2012competing} or propose alternative measures of influence \cite{cosley10-1,klemm2012measure}. 

Following the original proposal of Kitsak \emph{et al} \cite{kitsak2010identification}, Borge-Holthoefer and Moreno \cite{borge2012absence} studied rumor spreading dynamics to learn whether the $k$-core could predict authority or not. Surprisingly, their results indicate that a rumor's success $-$ measured as the number of individuals that learn about the rumor at the end of the spreading dynamics $-$ is topology-independent: no matter who in a network triggers the rumor, the final number of nodes who learn about it will be the same (given the same spreading parameters). Additionally, central nodes (those at the highest core levels) behave as firewalls, short-circuiting the capacity of the rumor to spread further. This theoretical prediction is clearly at odds with empirical evidence and points to a shortcoming of theoretical models that must be overcome.

The development of the Web $2.0$ and the growing popularity of online social networks have not only had a tremendous impact on our daily lives, but they also had the beneficial consequence of generating detailed data on social communication patterns, which can ultimately inspire and guide the development of more realistic models. In this paper we try to fill the gap between observations from real systems and theoretical predictions by introducing some simple modifications to models proposed previously \cite{moreno04efficiency,nekovee2007theory}. The resulting models are able to better approximate the behavior of users as observed in online social networks, in particular, the fact that there are influential nodes with larger diffusion capacities, an important feature not accounted for with current rumor spreading models.

Our analysis starts from the simple empirical observation \cite{barabasi2005origin,perra12-1,borge2012locating} that individuals display complex activity patterns both on and offline and, in particular, are not active around the clock. This fact has two possible interpretations. On one hand, users that are actually spreading the rumor are active only at specific times and only then they are able to participate in the diffusion process. On the other hand, an individual's choice of becoming active and participating in a specific information cascade can be seen as a demonstration of interest in the topic and his/her will to spread it. Inspired by these two interpretations, we derive two different rumor diffusion models. 

The first model incorporates the differences in the activity of the individuals responsible for the spreading of the rumor. Each spreader is assigned with a randomly chosen probability of being active at a given time. In this context, we study the effects of the heterogeneity \cite{perra12-1}  in the activation probability extracting values from three different probability distributions ranging from a uniform to a long-tailed one. In a more realistic version, following the idea that more active users, usually, also have a central role in the topology of the network, we relate the activity of each individual with its degree. 

The second model takes into account the fact that an individual could learn the rumor without actually spreading it further. This is for example what happens in most online social networks, in which followers receive pieces of information from those they are following and not always $-$ indeed, rarely $-$ they transmit the news further. We therefore introduce the possibility that a person that comes into contact with the rumor does not spread it anymore. This approach is complementary to the previous one, as we consider that it is the ignorants and not the spreaders who can be inactive. 

In the remaining of the paper,  we will show that even though these alternatives introduce only small and intuitive changes, they are able to shed light on the complex social mechanisms at work in real social systems and, at least qualitatively, reproduce the  heterogeneities observed in Twitter data. The rest of the manuscript is organized as follows: In the next section, we present a general framework for rumor spreading on networks while subsections \ref{sub1} and \ref{sub2} present the two modified models and the results of numerical simulations. Finally, we draw our conclusions in section \ref{sec3}.

%%%%%%%%%%%%%%%%%%%%%%%%%%%%%%%%%%%%%%%%%%%%%%
%%%%%%%%%%%%%%%%%%%%%%%%%%%%%%%%%%%%%%%%%%%%%%
\section{General modeling framework}
\label{sec:1}
In classical rumor spreading models on networks,  each of the $N$ nodes of a network can be in one of three possible states. A node holding a rumor and willing to transmit it is called a {\em spreader}. Nodes that are unaware of the update will be referred to as {\em ignorants}, while those that already know it but are not willing to spread it further are called {\em stiflers}. We denote the density of ignorants, spreaders, and stiflers at time $t$ as $i\left(t\right)$, $\rho\left(t\right)$ and $r\left(t\right)$, respectively, with $i\left(t\right) + \rho\left(t\right) + r\left(t\right) = 1$, $\forall t$. The spreading process takes place along the links connecting spreaders and ignorants. At each time step, spreaders contact all of their neighboring nodes. In the simplest case, whenever a spreader $j$ contacts a node $n$ that is ignorant, the latter will become a spreader with a fixed probability $\lambda$. Otherwise, if $n$ is already a spreader, the node $j$ will turn into a stifler with probability $\alpha$. Mathematically, the general model can be represented as:

\begin{eqnarray*}
I \stackrel{\lambda}{\longrightarrow} S \\
S \stackrel{\alpha}{\longrightarrow} R
\end{eqnarray*}
where the initial conditions are set such that  $i\left(0\right) = 1-1/N$, $\rho\left(0\right) = 1/N$ and $r\left(0\right)=0$. In addition, and without loss of generality, we set $\lambda = 1$ unless other values are explicitly stated.

\begin{figure}
  \centering
  \includegraphics[width=0.9\columnwidth,clip]{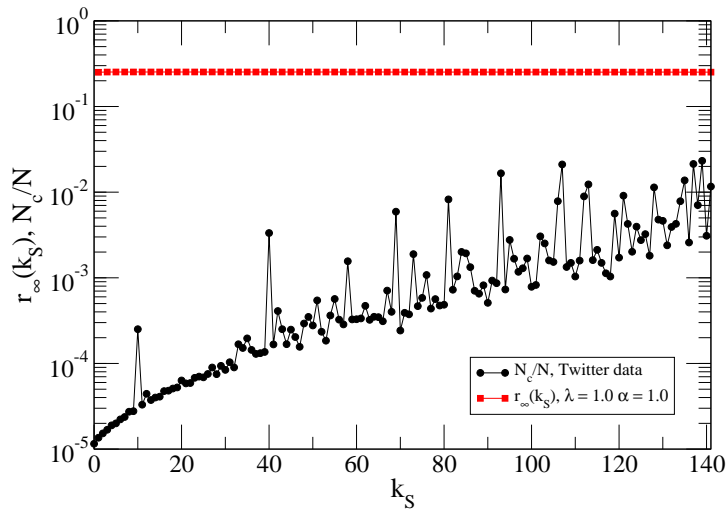}
  \caption{(color online) Density of stiflers at the end of the rumor spreading $r_{\infty}(k_{S})$ originated in a node that is part of the $k_{S}$ {\em k-core} for the general model of rumor spreading (\textit{squares}) and the empirical fraction, $N_c/N$, of users reached by cascades originated at nodes with {\em k-core} $k_S$ (\textit{circles}) as extracted by real Twitter data (for details see \cite{borge2012absence,gonzalez2011dynamics}). Numerical simulations were ran using the same empirical Twitter Follower network.}
  \label{fig1}
\end{figure}

For each alternative model presented, extensive numerical calculations have been carried out by simulating the dynamics of rumor propagation on top of a real-world Twitter following/follower network \cite{gonzalez2011dynamics}. From an initial scenario, in which all nodes belong to the ignorants class except the seed, we perform $S=10$ simulations. This is repeated for each node, i.e. every vertex of a network of $N$ nodes acts as the initial seed $S$ times, to obtain statistically significant results.  In this way, for each node $i$, we average the final density of stiflers in the network $r^{i}_{\infty}$. This quantity accounts for the spreading capacity of node $i$, which quantifies how deep the rumor penetrated the network when node $i$ was the initial seed:

\begin{equation}
r^{i}_{\infty} = \frac{1}{S}\sum_{m=1}^{S}r^{i,m}_{\infty}
\end{equation}
where $r^{i,m}_{\infty}$ represents the final density of stiflers for a particular run $m$ with origin at node $i$. With this information at hand for all nodes, we coarse-grain the individual $r^{i}_{\infty}$'s into classes of nodes according to their core number. Thus, $r_{\infty}(k_{S})$ represents the average stifler density for all runs with a seed with a $k_{S}$ core index:

\begin{equation}
r_{\infty}\left(k_{S}\right) = \sum_{i \in \Upsilon_{k_{S}}} \frac{r^{i}_{\infty}}{N_{k_{S}}}
\end{equation}
where $\Upsilon_{k_{S}}$ is the set of all $N_{k_{S}}$ nodes with $k_{S}$ values.

Figure \ref{fig1} shows the comparison between the values of $r_{\infty}\left(k_{S}\right)$ as obtained via the numerical simulations of the above rumor spreading model and the observed fraction of users $N_c/N$ reached  by cascades originated at nodes with core index $k_S$ obtained by analyzing Twitter usage data from the Spanish {\em Indignados} movement (see \cite{borge2012absence,gonzalez2011dynamics} for details on how data have been extracted and analyzed). The differences observed in this plot are striking. Even though the model ran on the exact same network, the theoretical prediction is completely insensitive to the value of the originating $k$-core, while in the empirical data there is a clear correlation between belonging to higher cores and larger numbers of  nodes reached by the cascade. This difference in behavior clearly shows that there is something fundamentally lacking in the theoretical model.

%%%%%%%%%%%%%%%%%%%%%%%%%%%%
\subsection{Model I: Human activity and temporal patterns}
\label{sub1}

We next consider the possibility that nodes are not always available to take part in a certain communication exchange. Each individual is active with a certain probability, $a_{i}$, affecting his/her behavior as a spreader. Thus, on top of the constraints of the basic framework presented above, we assume that a spreader only attempts to spread the rumor when it is active. As a consequence, the transition from the class of ignorants to the class of spreaders happens less often.

It is worth mentioning that as far as our model is concerned, the approach adopted is rooted in the observation that human activity patterns are mostly heterogeneous and therefore individuals are not always active \cite{perra12-1} nor is their activity distributed randomly over time \cite{barabasi2005origin,goncalves08-2,vazquez2007impact}. However, we assume that nodes in the network still have memory of who their potential neighbors are, and although not all the links of a given node were concurrently active, the set of available neighbors would be predefined by the underlying static (aggregated) topology. A more accurate description would require to consider that the topology is shaped by the activity of the nodes, so that the resulting time-varying networks are activity-driven \cite{perra12-1}. In the latter case, the interactions between the different classes of nodes in the system would still be activity-driven, but no memory of the static topology would be present, as the interaction structure is redefined at each time step. Whether or not both mechanisms lead to similar behavior is a matter that deserves further investigation.

On the other hand, note that being active or not has no effect on the rumor's recipients (ignorants). This mechanism is specific for asynchronous communication systems such as Twitter, FedEx,  email or SMS where information can be sent even without requiring the collaboration of the recipient. On the contrary, for synchronous systems, such as phone calls or Instant Messaging, that require both the source and the target of a message to be active at the same time, such a scheme would not suffice.  

Here we explore three possibilities for the activity distribution: (a) uniform, $P\left(a\right)\sim c$; (b) exponential, $P\left(a\right)\sim e^{-a/a_{c}}$; and (c) power-law, $P\left(a\right)\sim a^{-\gamma}$.
\begin{figure}
  \centering
  \includegraphics[width=\columnwidth,clip]{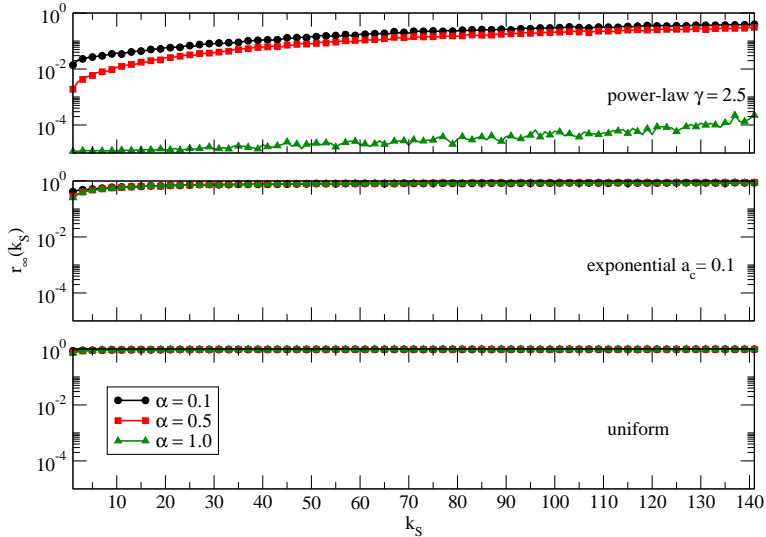}
  \caption{(color online) Density of stiflers at the end of the rumor spreading $r_{\infty}(k_{S})$ originated in a node of {\em k-core} $k_S$ in the activation probability model. Three different probability distribution functions are used: a power-law with exponent $\gamma = 2.5$ (upper panel),  exponential with $a_{c}=0.1$ (middle panel) and a uniform distribution (lower panel).}
  \label{fig2}
\end{figure}
Interestingly, these distributions yield completely different results. Figure \ref{fig2} illustrates this perfectly. The increase of heterogeneity in activity patterns moves the distribution of outbreak sizes,  $r_{\infty}\left(k_{S}\right)$, closer to empirical results, highlighting the fact that heterogeneity is a fundamental factor in real information spreading processes. A uniform activity distribution (lowest panel) completely flattens the spreading capabilities, no matter if nodes are in a topologically relevant region or not. This is in good agreement with \cite{borge2012absence}, the only difference being the time the system needs to reach a final state (that is, the probabilities delay the process significantly). An exponential distribution introduces some amount of asymmetries in the activity distribution, which slightly affects the spreading results (central panel). Finally, a power-law probability distribution introduces heterogeneity in the spreading success, the higher $k_{S}$, the higher the spreading capacity, just as it has been found empirically \cite{borge2012locating,gonzalez2011dynamics}.

The importance of the spreader-to-stifler rate is revealed in the heterogeneous scenario. $\alpha$ sets the timescale relevant to this process. For high values of $\alpha$, $\rho$ nodes quickly become stiflers and the rumor doesn't have the possibility of reaching a significant fraction of the population while  lower values of $\alpha$ easily allow for successful dissemination. Furthermore, it should be noted that we are assigning activity levels entirely at random, without any relation between topological features and activity probabilities. This means that a poorly connected node is just as likely to be highly active as a node with high degree. However, it has been seen previously~\cite{perra12-1} that activity distributions are correlated with the observed degree distribution. The simplest form of effectively implementing this correlation is to assign to node $i$ an activity probability $a_{i}=k_{i}/k_{max}$.

\begin{figure}
  \centering
  \includegraphics[width= \columnwidth,clip]{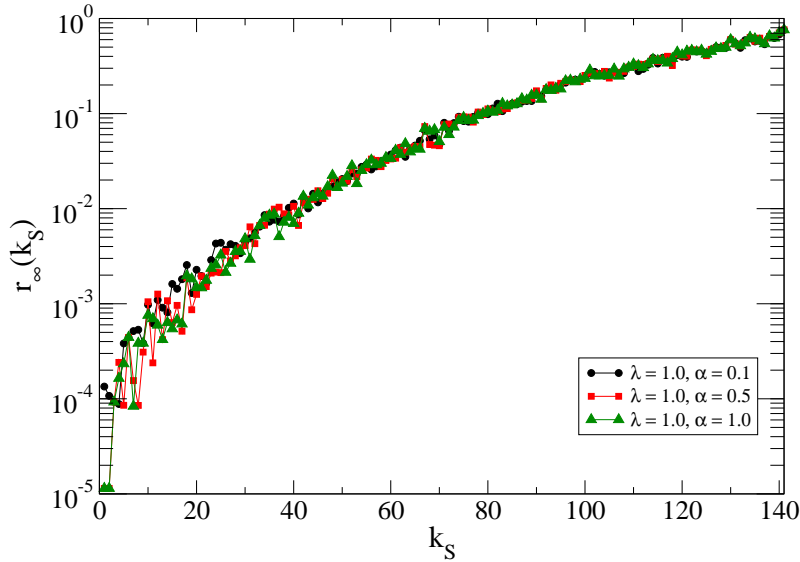}
  \caption{(color online) Fraction of stiflers at the end of the rumor spreading $r_{\infty}(k_{S})$ originated in a node of {\em k-core} $k_S$ when activation probabilities are proportional to nodes degree. Three different values of $\alpha$ are used, the underlying network is the empirical Twitter follower network from Ref. \cite{gonzalez2011dynamics}.}
  \label{fig3}
\end{figure}

Figure \ref{fig3} illustrates the results of this scenario. The great heterogeneity of the degree distribution is clearly reflected. Rumors triggered from low degree nodes (which necessarily have low $k_{S}$) die out soon, because the nodes they reach are almost never active. On the contrary, high degree nodes (which are more likely to belong to a high $k$ core) persistently forward messages, turning any rumor into system-wide knowledge. Note that spreading is almost identical regardless $\alpha$, in stark contrast to the upper panel of Figure \ref{fig2} (where $\alpha$ determines the shape of spreading) possibly indicating that higher level correlations also play an important role.

\begin{figure}
  \centering
  \includegraphics[width= \columnwidth,clip]{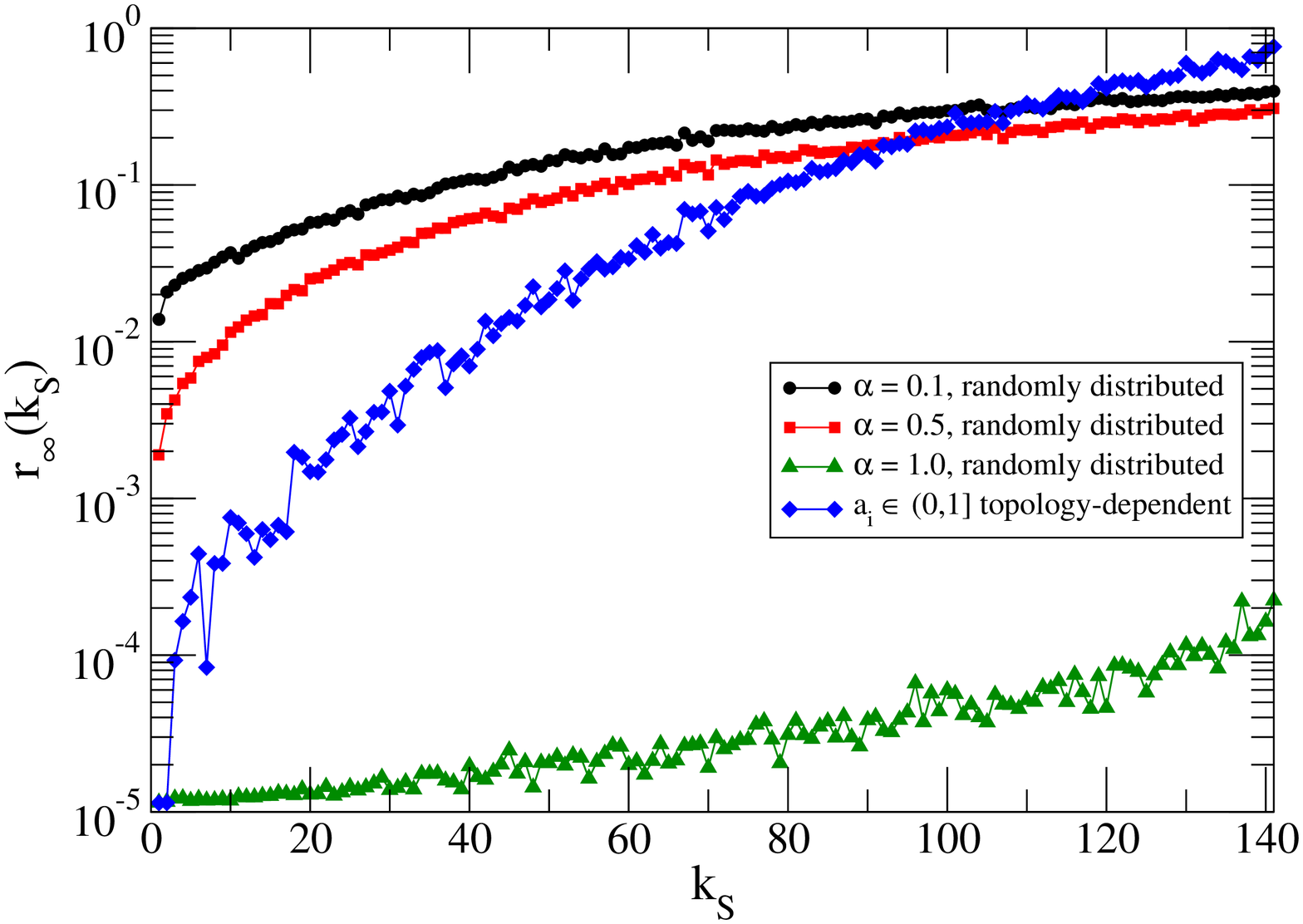}
  \caption{(color online)  Fraction of stiflers at the end of the rumor spreading $r_{\infty}(k_{S})$ originated in a node of {\em k-core} $k_S$ when the activation probability is proportional to nodes degree ({\em diamonds}) or randomly distributed ({\em squares}, {\em triangles} and {\em circles}). Network  topology and the other parameters are the same as in Fig.~\ref{fig3}.}
  \label{fig4}
\end{figure}

Figure \ref{fig4} shows the comparison between the topology-dependent and random distribution of the activation probabilities. In this case, all the curves have been obtained with a power-law activity distribution. However, in one of them (blue diamonds) the activity of each node is proportional to its degree ($a_{i} = k_{i}/k_{max}$), whereas in the other curves activation probabilities are assigned at random and  thus independently from the topological features of the nodes. As in Fig.~\ref{fig2} in the randomly distributed case the  spreading is highly affected by $\alpha$ meanwhile for the degree-dependent activation probabilities a substantial independence from $\alpha$ is present. 

%%%%%%%%%%%%%%%%%%%%%%%%%%%%
\subsection{Model II: Apathy}
\label{sub2}
The analysis of real data from online social networks demonstrates that most of the time users do not react to received messages \cite{borge2012locating}. One possible interpretation for this is that they have been informed of a rumor but chose not to spread it. This interpretation suggests another ingredient that might be missing from classical rumor spreading models and that might help bring them closer to reality: the possibility that an {\it ignorant} is apathetic and directly goes to the {\it stifler} status and does not participate further in the spreading dynamics.  As noted before, this kind of behavior is common in online social networks like Twitter, in which one receives messages that are rarely spread further. We incorporate this new element by introducing the probability, $p$, that an {\it ignorant} is interested in the topic and decides to diffuse it. In this scenario, when a {\it spreader} contacts an {\it ignorant}, the latter turns into a {\it spreader} with probability $\lambda p$ and into a {\it stifler} with probability $\left(1-p\right)\lambda$. The transitions allowed by our model are then:

\begin{eqnarray*}
I \stackrel{\lambda p}{\longrightarrow} S \\
S \stackrel{\alpha}{\longrightarrow} R\\
I \stackrel{\lambda (1-p)}{\longrightarrow} R
\end{eqnarray*}
It should be noted that Model II is a natural counterpart of Model I presented in section \ref{sub1}, in the sense that it also assigns activity probabilities to each node. The main difference is that this uniform probability $p$ is assigned to {\it ignorant} individuals and determines whether or not they choose to participate in the spreading process. A parallel can also be made to the case of epidemic spreading where a person become immune to a disease upon coming in contact with pathogen and before it is able to develop symptoms or spread it further.

The behavior of the system can be better understood analytically by writing the mean-field rate equations governing its time evolution in the homogeneous mixing approximation:

\begin{eqnarray}
\frac{\partial i\left(t\right)}{\partial t}= - \lambda p  \bar{k}  \rho\left(t\right)i\left(t\right) - \left(1-p\right)\lambda \bar{k}\rho\left(t\right) i\left(t\right),\label{eq:i} \\
\frac{\partial \rho\left(t\right)}{\partial t}=  \lambda p \bar{k} \rho\left(t\right)i\left(t\right)  -\alpha \bar{k} \rho\left(t\right)\left(\rho\left(t\right)+r\left(t\right)\right), \label{eq:s}\\ 
\frac{\partial r\left(t\right)}{\partial t}= \alpha \bar{k} \rho\left(t\right)\left(\rho\left(t\right)+r\left(t\right)\right) + \left(1-p\right)\lambda \bar{k}\rho\left(t\right) i\left(t\right), \label{eq:r}
\end{eqnarray}
with the initial conditions $i\left(0\right)=1-1/N, \rho\left(0\right)= 1/N$, $r\left(0\right)=0$ and  where $\bar{k}$ represents the number of contacts each spreader has per unit time. The first term in the right side of Eq.~\ref{eq:i} accounts for the density of ignorants that turn into spreaders after an interaction whereas the second term model the ignorant to stifler transition with probability $\left(1-p\right)\lambda$. 

Recalling that $i\left(t\right)+\rho\left(t\right)+r\left(t\right)=1$  we can study the system of Eqs.~\ref{eq:i}-\ref{eq:r} analytically in the infinite-time limit $\rho({\infty})=0$, obtaining:
\begin{equation}
r_{\infty} = 1-e^{- (1+ \frac{\lambda}{\alpha} p ) r_{\infty}}.
\label{eq:rinf}
\end{equation}
\begin{figure}
  \centering
  \includegraphics[width= \columnwidth,clip]{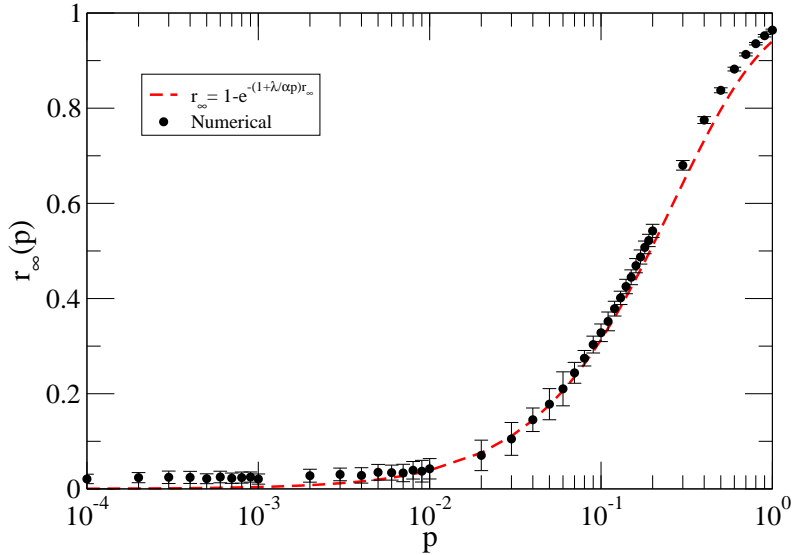}
  \caption{(color online) Fraction of stiflers at the end of the rumor spreading $r_{\infty}$ for different values of $p$  in comparison with the theoretical prediction of Eq.~\ref{eq:rinf}. Numerical results are the average over $10^{3}$ stochastic runs. In both cases $\lambda= 1.0$, $\alpha = 0.5$ and $\bar{k}  = 4$. }
  \label{fig5}
\end{figure}
The average total stifler density $r_{\infty}$, for various values of $p$, obtained by numerically solving this transcendental equation is shown in Figure \ref{fig5}. We also performed a series of Monte-Carlo (MC) simulations in the homogenous mixing limit.  At $t=0$ the entire population is {\it ignorant} with only a small fraction ($ \simeq 1/N $) being spreaders. At each time step, each spreader contacts $\bar{k}$ individuals chosen at random from the entire population. If the chosen individual is an ignorant it will become a spreader with probability $\lambda p$ or directly move to stifler status with $\left(1-p\right)\lambda$. Otherwise, when a spreader comes in contact with a stifler or another spreader it turns into a stifler with probability $\alpha$. When the spreading process reaches the absorbing state $\rho\left(t\right) = 0$ the final density of stiflers is recorded. The simulation results are also plotted in Figure~\ref{fig5} for comparison with the analytical solution. The agreement between the two approaches is striking and serves as a confirmation that we are not missing any fundamental ingredients in our analyses.

 \begin{figure}
  \centering
  \includegraphics[width= \columnwidth,clip]{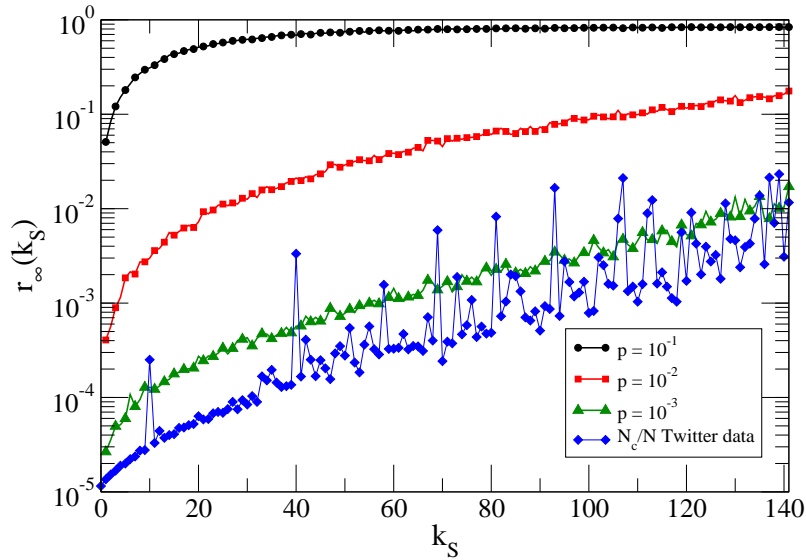}
  \caption{(color online) Fraction of stiflers at the end of the rumor spreading $r_{\infty}(k_{S})$ originated in a node of {\em k-core} $k_S$ in the ignorant-to-stifler transition model for different $p$ values. For each value of $p$ both the average and the maximum value of numerical simulations is  presented. Network topology and the other parameters are the same as in Fig.~\ref{fig3}.}
  \label{fig6}
\end{figure}

Although the addition of a constant $p$ parameter (any node is assigned the same $p$) is a crude approximation to the interest that {\it ignorants} might have in becoming spreaders, it has profound implications for the system dynamics when compared to the standard setup. Figure \ref{fig6} shows the behavior of the system with the inclusion of the new rule, with fixed $\lambda = 1$, $\alpha = 0.5$ for different $p$ values. As for the power-law activity distribution in model $I$ and in real data \cite{borge2012locating,gonzalez2011dynamics,perra12-1} a strong correlation between the {\em k-core} of the seed and the final outcome of the spreading is observed. In particular, and although we have made no efforts toward fitting this value, it is clear that for a very low probability ($p = 10^{-3}$) we already have a close-to-real behavior. Although this value might seem low (only one in a thousand contacted individuals do forward the rumor),  one must consider that this is the probability that one individual will choose to participate in any of rumors he observes. It is well known that most Twitter users commonly follow on the order of hundreds of other individuals so that the number of pieces of content  they are exposed to daily can easily be on the order of thousands or tens of thousands of which they are only able, or willing, to participate in a few. %Values of this order have also been seen previously in \cite{borge2012locating}. 
 
\section{Conclusions}
\label{sec3}
Online social networks are becoming increasingly central in our lives as they come to permeate our daily activity. It then comes as no surprise that they have been welcome by mass social movements around the world as unique platforms for the diffusion of new ideas and even for the coordination of large numbers of individuals. Understanding the forces that drive the behavior of individuals interacting in these networks is then one of the great challenges for science in the next years. 

One interesting aspect is  how ideas are shared between individuals and what are the conditions that allow for a large dissemination of them. In this context several works studied how a rumor can spread in a population of {\it ignorant} individuals but, due to the changes in the way in which these tools allow us to communicate, most of those works cannot catch the details of rumor dynamics on such large scale social systems.

Driven by data from a microblogging online platform we propose two modifications to classical rumor spreading models that are able to qualitatively reproduce the observed differences in the number of individuals reached by the rumor when the seed is located in the most connected circles of the network or in its periphery.  The models we present are based on the observation that individuals, both {\it spreaders} and {\it ignorants}, are not always active in the network. Each model then implements a different effective mechanism that is consistent with this fact: Model I assigns activity probabilities to each node and allows for spreading to occur only when a spreader node is active while Model II assumes that each node has a finite probability of being interested in spreading each specific rumor and would otherwise chose not to participate in the diffusion process. Both variations have proved effective in bringing the classical model one step closer to reality.

In the case of Model I, numerical results highlight that the more heterogeneous the patterns of activation are the more faithfully we are able to imitate real data. Moreover, if, in a second approximation, we relate the activity of a node with its degree (as higher degrees are commonly correlated with high levels of activity in the network) we also observe a substantial independence of the results from stifler transition ratio, $\alpha$. For Model II, we were also able to give an analytical expression for the final density of stiflers in the system. Interestingly, the analysis of the numerical simulations suggests that close-to-real results are obtained when the probability for an ignorant to be interested in the rumor is very low; another feature also observed in real social networks. 

The results presented in this paper clearly evidence that classical rumor spreading models are severely short on their ability to effectively approximate reality. We have shown that even small, empirically based, modifications can significantly increase their level of realism. In particular, our results shed some light on the interplay between technology and human interactions that are at the origin of some of the complex behaviors we observe daily. With this work we have taken a significant first step in paving the way toward a deeper understanding of how ideas spread through our online and offline social networks and help shape current events and society as a whole.

\begin{acknowledgements}

The authors would like to thank A. Rivero for the help in collecting Twitter data and useful discussions.
This work has been partially supported by MICINN through Grants No. FIS2008-01240, FIS2009-13364-C02-01 and No. FIS2011-25167, and by the Government of  Arag\'on (DGA)  through a grant to FENOL group.
\end{acknowledgements}

\bibliographystyle{spmpsci}      % mathematics and physical sciences
\bibliography{biblio}   % name your BibTeX data base

\end{document}